\documentclass[conference]{IEEEtran}
\IEEEoverridecommandlockouts
\usepackage{cite}
\usepackage{amsmath,amssymb,amsfonts}
\usepackage{algorithmic}
\usepackage{graphicx}
\usepackage{textcomp}
\usepackage{xcolor}

\def\BibTeX{{\rm B\kern-.05em{\sc i\kern-.025em b}\kern-.08em
    T\kern-.1667em\lower.7ex\hbox{E}\kern-.125emX}}
\begin{document}

\title{The Effect of Noise on the Performance
of Variational Algorithms for Quantum Chemistry
}

\author{\IEEEauthorblockN{Waheeda Saib}
\IEEEauthorblockA{ \textit{IBM Quantum, IBM Research-Africa}\\
Johannesburg, South Africa \\
wsaib@za.ibm.com}
\and
\IEEEauthorblockN{Petros Wallden}
\IEEEauthorblockA{\textit{University of Edinburgh}\\
Edinburgh, Scotland \\
petros.wallden@ed.ac.uk}
\and
\IEEEauthorblockN{Ismail Akhalwaya}
\IEEEauthorblockA{
\textit{IBM Research-Africa, University of the Witwatersrand}\\
Johannesburg, South Africa \\
ismaila@za.ibm.com}
}

\maketitle

\thispagestyle{plain}
\pagestyle{plain}

\begin{abstract}
Variational quantum algorithms are suitable for use on noisy quantum systems. One of the most important use-cases is the quantum simulation of materials, using the variational quantum eigensolver (VQE). To optimize VQE performance, a suitable parameterized quantum circuit (ansatz) must be selected. We investigate a class of ansatze that incorporates knowledge of the quantum hardware, namely the hardware efficient ansatze. The performance of hardware efficient ansatze is affected differently by noise, and our goal is to study the effect of noise on evaluating which ansatz gives more accurate results in practice. First, we study the effect of noise on the different hardware efficient ansatze by benchmarking and ranking the performance of each ansatz family (i) on a chemistry application using VQE and (ii) by the recently established metric of “expressibility”. The results demonstrate the ranking of optimal circuits does not remain constant in the presence of noise. Second, we evaluate the suitability of the expressibility measure in this context by performing a correlation study between expressibility and the performance of the same circuits on a chemistry application using VQE. Our simulations reveal a weak correlation and therefore demonstrate that expressibility is not an adequate measure to quantify the effectiveness of parameterized quantum circuits for quantum chemistry. Third, we evaluate the effect of different quantum device noise models on the ordering of which ansatz family is best. Interestingly, we see that to decide which ansatz is optimal for use, one needs to consider the specific hardware used even within the same family of quantum hardware.
\end{abstract}


\section{Introduction}

One of the early inspirations that led to the development of quantum computation was Feynman's view that quantum systems are better in simulating other quantum systems \cite{P8}. More recently, precisely for this reason (fundamental quantum nature),  quantum chemistry has been identified as one of the most promising fields to demonstrate useful quantum computational advantage \cite{P2}. Examples that could benefit from such endeavors is the comprehensive analysis of chemical systems, their properties and reaction rates, which may yield advancements on currently intractable problems such as designing new drugs and catalysts to aid the biological process of nitrogen fixation \cite{P1,P2,P3}.

Existing and near term quantum hardware have several shortcomings such as limited qubit number, qubit connectivity and importantly, gate errors and decoherence. It is therefore impossible to run quantum error correction on them and one needs to focus on methods and algorithms that directly mitigate the effects of noise. Variational quantum algorithms (VQA), a class of hybrid quantum classical algorithms, are designed to break the problem in question to a part that uses the quantumness (and is in principle intractable classically), while delegate the remaining part to classical computers. The Variational Quantum Eigensolver (VQE) \cite{P12,P13,P11} was one of the first such algorithm and is designed to approximate the ground state energy of a Hamiltonian. VQE is naturally well suited to recover the ground state energy of small molecules \cite{P1}.
A crucial component of VQE, is the selection of a suitable parameterized circuit or ansatz that represents the solution space of the problem. The two types of ansatze that exist in literature, are the ``problem specific'' ansatz that uses the details of the problem one attempts to solve (e.g. the coupled cluster method \cite{P1,P2} for chemistry problems), and the hardware efficient ansatz \cite{P1,P2,P11} that chooses a family of states that is ``easy'' for the hardware one uses, but can be used for any problem. In this contribution we focus on the latter (hardware efficient) and when we refer to ansatz in the remaining text it is understood as hardware efficient ansatz. 

Many different ansatze exist. Interestingly the effects of (specific) noise affect the performance of each of these ansatze within a VQE algorithm differently. 
In this work, we want to 
assess the effect of noise on the decision of which ansatz family gives more accurate results in practice. First, we study the effect of noise on the different ansatze by benchmarking and ranking the performance of each ansatz family by expressibility as well as by its performance on a chemistry application using VQE. The simulated results demonstrate the ranking of optimal circuits does not remain constant in the presence of noise. Second, the expressibility measure is evaluated by performing a correlation study between expressibility and the performance of the circuits on a quantum chemistry problem using VQE for ideal and noisy quantum conditions. The results reveal, expressibility displays a weak correlation and is not an appropriate measure to quantify the efficacy of parameterized quantum circuits in the quantum chemistry domain. The underlying reason for this weak correlation is examined by performing non-uniform parameter sampling, that shows a significant increase in expressibility by simply changing the parameter sampling methodology for a given circuit. 
Third, the effect of noise models based on different quantum devices are examined on the ranking of the best performing circuits. The simulations demonstrate, the order of the optimal circuits to use for a given problem does not remain constant as the performance of the circuits vary with different noise levels, even within the same family of quantum hardware.

This work facilitates a deeper understanding of how quantum device noise impacts the decision of the ansatz family to use even among different devices that have the same quantum computing hardware and provides new insight into the applicability of the, widely used, expressibility metric.\\

\section{Related works} The review \cite{P2} highlights that studies comparing the effectiveness of ansatz have been performed for small chemistry problems, but the focus is different from our work.  The expressibility measure was developed to offer insight on the selection of parameterized circuit for a given task or algorithm \cite{P4}. Expressibility was used to understand and evaluate the capability of quantum circuits employed in quantum chemistry and quantum machine learning literature \cite{P4}.  Recent literature, used expressibility to benchmark parameterized quantum circuits for quantum machine learning applications \cite{P6} and to assess the alternating layered ansatz for a quantum chemistry application \cite{P5}, under ideal quantum conditions. In our work, we build on this literature to assess the ability of the expressibility measure to identify effective ansatze for a chemistry problem under noise-free and noisy quantum conditions. 

Recent studies exploring the effect of noise on variational quantum algorithms, employed different noise models, types of noise and noise levels to assess the VQE performance using the hardware efficient ansatz \cite{P28,P29}. These studies only explore two specific hardware efficient circuit layouts with different parameterized gates and circuit layers, while the effect of noise is not as well studied for quantum chemistry. We complete this analysis by considering a wider class of ansatze and examine the effects it has for the specific application of quantum chemistry using VQE. 

\section{Methods}

\subsection{Variational Quantum Eigensolver}

The VQE algorithm 
is used to approximate the ground state energy of quantum systems 
\cite{P1,P11}. It 
consists of four main tasks. First, the quantum computer is used to prepare an initial trial quantum state \(|\psi(\theta)\rangle\) by assigning a set of parameters $\theta$ to parameterized gates on the quantum circuit \cite{P1}.

Second, the associated energy of each trial state is estimated by using the Hamiltonian averaging method. In this step the Hamiltonian is decomposed into Pauli terms \( P_\alpha = {\sigma_1}^{\alpha_1} \otimes {\sigma_2}^{\alpha_2} \otimes ... {\sigma_N}^{\alpha_N} \), with N representing the number of qubits and each Pauli operator in $P_\alpha$ representing one of $I,X,Y,Z$ such that the Hamilton is represented as a linear combination of Pauli terms

\begin{equation}
H = \sum_\alpha h_\alpha P_\alpha
\end{equation}
where $h_\alpha$ denotes a real scalar coefficient and $\alpha$ represents the term in the Hamiltonian \cite{P1,P2}. The expectation value of each Pauli term in the problem Hamiltonian is estimated by repeatedly running the trial state preparation and measurement steps (where the corresponding Pauli observable is measured). 

Third, On the classical computer the expectation values of $P_\alpha$ are summed with corresponding weights $h_\alpha$ to calculate the total energy or cost function \cite{P1}.
\begin{equation} 
E(\theta) = \langle{\Psi(\theta)} |H|{\Psi(\theta)}\rangle = \sum_\alpha h_\alpha \langle{\Psi(\theta)} |P_\alpha|{\Psi(\theta)}\rangle \end{equation}

Fourth, the classical optimization method is used to generate a new set of parameters $\theta$ that minimize the total energy or cost function \cite{P46}. The new parameters are fed into the parameterized gates on the quantum circuit to prepare a new trial state \cite{P2}. The aim being to find the quantum state that has the smallest energy and thus to get an approximation of the actual ground state energy. 
This process is repeated until the optimal parameters are found that generate the minimal energy within an acceptable accuracy \cite{P1}. There are several components of the VQE algorithm that may be improved to increase the performance of the algorithm. The components that may be modified are the trial state or ansatz, classical optimization methods, mappings from fermions to qubits, circuit optimization and error mitigation techniques \cite{P1}. In this contribution we focus on the choice of ansatz.

\subsubsection{Hardware Efficient Ansatz}
The hardware efficient ansatz (HEA) is a parameterized quantum circuit that consists of a series of parameterized single qubit gates and two-qubit entangling gates that are tailored to the quantum gate set of the available quantum hardware \cite{P1}. 
The main benefit of hardware efficient ansatz is that it is domain agnostic and versatile, as it has been demonstrated to perform well on quantum chemistry\cite{P11} and quantum machine learning problems\cite{P18}.

\subsubsection{Simultaneous Perturbation Stochastic Approximation}
The classical optimization method, simultaneous perturbation stochastic approximation (SPSA) has been successfully applied to noisy quantum chemistry problems, that find the ground state energy of small molecules \cite{P11}. The SPSA method reduces the sampling overhead by approximating the gradient using two energy estimates irrespective of the parameter space dimension \cite{P11}.

\subsection{Expressibility}

Expressibility is defined as the ability of a parameterized quantum circuit (PQC) to produce quantum states that represent the Hilbert space well \cite{P4}. The expressibility measure has been proposed as a new approach to benchmark the capability of parameterized quantum circuits \cite{P4}. A principled approach to quantify expressibility is outlined, by statistically computing numerical simulations of the parameterized quantum circuit \cite{P4}. The method to estimate expressibility, compares the probability distribution of quantum state fidelities computed from a set of quantum states sampled from the PQC to the uniform distribution of quantum fidelities from the ensemble of Haar random states \cite{P4}. 
The expressibility method may be broken down into four main steps. First, the parameterized quantum circuit is selected and initialized with parameters represented by $\theta$. Second, the distribution of quantum fidelities for the PQC are computed by uniformly sampling pairs of parameters and their associated quantum states. Quantum states are simulated by applying parameter vectors to the quantum circuit, thereafter the fidelity is computed by obtaining the squared overlap \(F = |\langle \psi_\theta|\psi_\phi \rangle |^2\) of two quantum states \cite{P4}. This experiment is repeated for $M$ samples to create a probability distribution of fidelities from the PQC $\hat{P}_{PQC}(F;\theta)$. Third, the analytical method to compute the probability density function of fidelities for the ensemble of Haar random states is expressed as $P_{Haar}(F) = (N-1)(1-F)^{N-2}$ where N is the dimension of the Hilbert space and F represents the fidelity \cite{P4,P14}. Fourth, the two probability distributions are discretized with a histogram and used to estimate the Kullback-Leibler (KL) divergence \cite{P22}, which is measured by taking the difference between the probability distribution of fidelities for the PQC from the ensemble of Haar random states to attain the value for expressibility 
\begin{equation} 
Expr = D_{KL}(\hat{P}_{PQC}(F;\theta) || P_{Haar}(F))
\end{equation} 
Expressibility can hence be defined as the amount of information lost when estimating the probability distribution of state fidelities from a PQC using the ensemble of Haar random states fidelities \cite{P4}. It follows that the smaller the value of expressibility, the closer to true random states (and thus the more expressible) is the parameterized circuit family.





\subsection{Noise model}

The IBM Quantum Experience provides access to several quantum processors with different qubit number, quantum volume, and noise levels \cite{P28}. Since the fundamental performance of the quantum processor is characterized by a set of error parameters namely, average gate error for one and two-qubit gates, gate duration, one qubit measurement error probabilities, T1 and T2 decoherence time for every qubit. The basic noise model of a quantum device may be constructed from these backend error parameters using Qiskit and implemented using Qiskit Aer, a high performance simulator, to simulate noisy quantum conditions \cite{P30}.

In comparison with noisy quantum devices the basic noise model makes several assumptions. The gate and measurement errors are assumed to be local and Markovian, that is cross-talk and leakage errors are excluded \cite{P30}. Gate errors are approximated by incoherent noise processes and modeled by a combination of depolarizing and thermal relaxation error \cite{P30}. The single qubit gate errors are modeled by single qubit depolarizing error and single qubit thermal relaxation error characterized by T1, T2 and gate duration \cite{P23}. Two-qubit gates consist of two-qubit depolarizing error followed by single qubit thermal relaxation errors for both gate qubits \cite{P23}. The strength of the depolarizing error is selected such that the gate error represents the average gate error from benchmarking experiments on the IBM quantum processors \cite{P30}.

\section{Experimental design and setup}
In this study, we explore different variations of the hardware efficient ansatz to assess the effect of noise on the decision of ansatz family to use for a quantum chemistry application. 


The hardware optimized circuit designs, outlined in Appendix, consist of parameterized single and two-qubit gates with a circuit depth of one and a circuit width of four qubits. The ideal and noisy quantum simulations are performed using the qasm simulator. The IBM Quantum Experience platform was used to generate circuit visualizations and access quantum device simulators with different noise models. The Qiskit framework \cite{P23} was used to implement our simulations. The experiments performed under noisy quantum conditions employed the 5 qubit IBMQX2 quantum device noise model. The experiments are performed on ideal and noisy IBM quantum simulators to understand and compare the performance of the methods under realistic noisy conditions. 

\subsection{Chemistry problem}
The electronic structure problem is concerned with finding the ground state energy of a molecular or chemical system \cite{P2}. We restrict ourselves to 
the hydrogen H$_2$ molecule, since it is a 
problem that has known results. We use it 
to benchmark and evaluate the performance of the hardware efficient circuits using the VQE algorithm, under noise-free and noisy quantum conditions.

\subsection{Mapping the chemistry problem to quantum computation}
To map a molecule to qubits requires few assumptions. 
The Born-Oppenheimer approximation is applied to simplify the molecular Hamiltonian of hydrogen by fixing the nuclei positions, since the nuclei are much larger than the electrons, they may be regarded as stationary classical particles \cite{P46}. The resulting Hamiltonian 
takes the fixed nuclei positions as parameters. 
The Hamiltonian is converted to second quantization using a basis set, STO-3G, which is a set of distinct functions, that approximately describes the spin-orbitals of a molecule \cite{P2}. The Hamiltonian in second quantization is expressed in terms of creation and annihilation operators 
\cite{P2}. Finally, we map the creation and annihilation operators of the Hamiltonian, to qubit operators using the Jordan-Wigner transformation \cite{P2}, to attain a four qubit system for H$_2$ \cite{P46}.

\subsection{The effect of noise on circuit ranking}
We compute and rank the performance of each ansatz using (i) the expressibility measure and (ii) by comparing the solution found using VQE with the (known in this case) ground state energy of a chemistry application.

\subsubsection{VQE Experiments}

Solving the electronic structure problem using the VQE algorithm appears a promising approach to useful quantum computational advantage \cite{P2,P12,P13}, but has also been used as a quantum chemistry simulation benchmark for near term quantum computers \cite{P25}. Here we use VQE to benchmark the hardware efficient circuits under noise-free and noisy quantum conditions.


The molecular system of hydrogen H$_2$ is transformed into the molecular Hamiltonian using the Qiskit Nature driver for the PySCF library in the STO-3G basis \cite{P23}. The PySCF library, is an open source computational chemistry library \cite{P47}. 
The VQE algorithm is implemented using the VQE standard established in \cite{P11}. 
The classical optimizer applied to the VQE algorithm, as established in \cite{P11}, is the SPSA optimizer. Since SPSA reduces the number of gradient computations required per iteration and performs well for fermionic optimization problems, this method is suitable for variational quantum algorithms applied to quantum chemistry \cite{P15}. The SPSA optimizer was employed for a maximum number of 200 iterations, to reduce the run time of the experiment. The performance metric used to assess the accuracy of the VQE simulations is the energy difference, which is the difference between the minimum energy reached by the simulation and the actual ground state energy of hydrogen H$_2$. The performance of the hardware optimized circuits applied to VQE are presented in Table~\ref{table:expr_vqe} and Table~\ref{table:expr_vqe_noise} for ideal and noisy quantum simulations.


\subsubsection{Expressibility Experiments}

The procedure to estimate expressibility is performed by first, selecting a parameterized circuit that may be initialized by $\theta$. Second, we uniformly sample two parameters that are applied to the parameterized circuit to generate the quantum states. Third, the two quantum states are used to attain the fidelity by computing the squared overlap between the states. In the ideal simulation the quantum states are pure, whereas in the noisy simulations the quantum states are mixed which requires further computation to attain the average of the quantum states, before the fidelity is obtained. This process is repeated 5000 times to attain a distribution of quantum state fidelities for each circuit in the Appendix. Fourth, the distribution of Haar random state fidelities, is analytically computed using $P_{Haar}(F) = (N-1)(1-F)^{N-2}$ where N represents the dimensionality of the Hilbert space and F is the fidelity. Fifth, the fidelity distributions from the parameterized circuit and the Haar random states are discretized using a histogram with a bin size of 75. A sample size of 5000 and a bin size of 75 were selected in order to reproduce the experiments and attain agreement for the hardware efficient circuit in \cite{P4}. Sixth, the expressibility measure is computed by taking the KL divergence between the probability distribution of state fidelities generated from the parameterized circuit and the Haar random states. The expressibility for the circuits are estimated and shown in the Appendix at Table~\ref{table:expr} and Table~\ref{table:expr_noise} for ideal and noisy quantum simulations.

\subsubsection{Variation of two-qubit gates}
The different ansatze used in this benchmark study consist of single and two-qubit entangling gates. The two-qubit gates used are the controlled-Z (CZ) and controlled-NOT (CX) gates. 
We use three fixed circuit designs, but we vary the type of two-qubit gates used. For example circuit 1 has only CX two-qubit gates, circuit 2 only CZ two-qubit gates, while circuits 3 to 8 have a mixture of CX and CZ gates.
The expressibility and VQE results of the circuits with different types and arrangements of two-qubit gates may be viewed in Table~\ref{table:expr_vqe} and Table~\ref{table:expr_vqe_noise} for ideal and noisy quantum simulations.

\subsubsection{Evaluation of the Expressibility measure} 
To assess the strength of the expressibility measure to estimate the capability of parameterized quantum circuits, we statistically analyzed the correlation between expressibility and the performance of hardware efficient circuits on a chemistry application using the VQE algorithm. The Pearson correlation coefficient and scatter plot diagrams are used to assess and visualize the relationship between expressibility and the energy difference. The Pearson correlation coefficient measures the strength of the relationship between two quantitative variables, assuming the relationship is linear \cite{P26}. Pandas, a python library for data analysis and statistics \cite{P27}, is used to compute the Pearson correlation coefficient between the expressibility and energy difference values for the hardware efficient circuits in the Appendix. The scatter plot diagrams were generated to show the relationship between the expressibility and energy difference values for each circuit in the Appendix. 
The scatter plot diagrams produced for ideal and noisy quantum simulations may be viewed in Fig.~\ref{fig:expr_scatterplot} and Fig.~\ref{fig:expr_scatterplot_noise}. 

We further assess whether the expressibility of a circuit remains constant or can be changed simply by changing the parameter sampling method used. This experiment is performed by using non-uniform parameter sampling instead of uniform sampling of the parameter space when estimating the expressibility measure of a fixed circuit. The
expressibility computed for the fixed circuit shown in Fig.~\ref{fig:circb_vis}, using non-uniform parameter sampling, is visualized in Fig.~\ref{fig:expr_sampling}.

\subsection{The effect of different noise models on circuit ranking}

To examine the effect of different noise models on the hardware optimized circuits, we ranked and evaluated the performance of the circuits applied to a chemistry problem using VQE for different IBM Quantum device noise models. Since depolarizing noise has been shown to accumulate with increased circuit depth \cite{P28}, the circuits under investigation have been fixed with a circuit depth of one. The experiment, utilizes the noise model from existing IBM quantum processors. 
The results 
demonstrating the variation in circuit ranking for different noise models are shown in Fig.~\ref{fig:noise_model_rank}.

\section{Results}


\subsection{VQE performance observations}
We compute and assess the performance of different ansatze when used within VQE to estimate the ground state energy of H$_2$. 
The results 
are summarized 
in Table~\ref{table:expr_vqe} for ideal and Table~\ref{table:expr_vqe_noise} for noisy quantum simulations. The tables contain information on the circuit ID as defined  
in the Appendix, the two-qubit gates used, the expressibility of the circuit, the computed ground state energy of hydrogen and the energy difference between the computed energy and the (known in this case) true ground state energy measured in hartree units. 
The true ground state energy of hydrogen is -1.1373 hartree. The estimated ground state energy, and the corresponding energy difference, for a given ansatz is computed from the energy of the circuit with the optimal parameters, i.e. the parameters that give the lowest energy for our problem. 
We rank the circuits by ascending values of the computed ground state energy and energy difference, that is presented in descending order from the best to the worst performing circuits, shown by the circuit ID column in Table~\ref{table:expr_vqe} for ideal and Table~\ref{table:expr_vqe_noise} for noisy quantum simulations. This allows us to monitor both the performance and to compare the correlation of performance of an ansatz with its expressibility.  

\begin{table}[t]
\caption{Ranking of ansatze by VQE simulation of hydrogen on ideal quantum simulator}
\begin{center}
\begin{tabular}{ |p{0.8cm}|p{2cm}|p{1.5cm}|p{1.25cm}|p{1.2cm}| } 
\hline
Circuit ID & Gates & Expressibility & Ground state energy & Energy diff. from correct\\
\hline
9 & CX & 0.648 & -1.1211 & 0.0161 \\ 
3 & CX,CZ,CX,CZ & 0.229 & -1.1205 & 0.0167 \\ 
7 & CZ,CX,CX,CZ & 0.224 & -1.1201 & 0.0171 \\ 
8 & CZ,CZ,CX,CX & 0.240 & -1.1194 & 0.0178 \\ 
1 & CX & 0.224 & -1.1192 & 0.0180 \\ 
12 & CX & 0.020 & -1.1190 & 0.0182 \\ 
11 & CZ & 0.027 & -1.1184 & 0.0188 \\ 
5 & CX,CX,CZ,CZ & 0.205 & -1.1179 & 0.0193 \\ 
4 & CX,CZ,CZ,CX & 0.234 & -1.1174 & 0.0198 \\ 
6 & CZ,CX,CZ,CX & 0.202 & -1.1155 & 0.0218 \\ 
2 & CZ & 0.226 & -1.1088 & 0.0284 \\ 
10 & CZ & 0.691 & -0.5276 & 0.6096 \\ 

\hline
\end{tabular}
\end{center}
\label{table:expr_vqe}
\end{table}


\noindent\emph{Ideal simulations.} In Table~\ref{table:expr_vqe} we observe that circuits 9 and 3 perform best (smallest energy difference -- 0.0161 and 0.0167 hartree respectively). Circuits 10 and 2 are the worst performing circuits (0.6096 and 0.0284 hartree). While in general we see that circuits with CX gates perform better than similar ones with CZ gates, interestingly we note that circuits 3, 7 and 8, that have mixed arrangements of the CX and CZ gates, perform better than circuit 1 that has only CX gates. 



\begin{table}[b]
\caption{Ranking of ansatze by VQE simulation of hydrogen on noisy quantum simulator}
\begin{center}
\begin{tabular}{ |p{0.8cm}|p{2cm}|p{1.5cm}|p{1.25cm}|p{1.2cm}| } 
\hline
Circuit ID & Gates & Expressibility & Ground state energy & Energy diff. from correct\\
\hline
2 & CZ & 0.673 & -0.9283 & 0.2089 \\ 
4 & CX,CZ,CZ,CX & 0.715 & -0.9263 & 0.2109 \\ 
7 & CZ,CX,CX,CZ & 0.832 & -0.9186 & 0.2186 \\ 
5 & CX,CX,CZ,CZ & 0.885 & 0.9104 & 0.2268 \\ 
12 & CX & 0.946 & -0.9084 & 0.2288 \\ 
11 & CZ & 1.459 & -0.9001 & 0.2371 \\ 
8 & CZ,CZ,CX,CX & 0.747 & -0.8967 & 0.2405 \\ 
3 & CX,CZ,CX,CZ & 0.731 & -0.8826 & 0.2546 \\ 
6 & CZ,CX,CZ,CX & 0.899 & -0.8805 & 0.2567 \\ 
1 & CX & 0.698 & -0.8718 & 0.2655 \\ 
9 & CX & 0.835 & -0.8397 & 0.2975 \\ 
10 & CZ & 2.668 & -0.3526 & 0.7846 \\ 

\hline
\end{tabular}
\end{center}
\label{table:expr_vqe_noise}
\end{table}

\noindent\emph{Noisy simulations.}
In Table~\ref{table:expr_vqe_noise} we observe that circuits 2 and 4 perform best in a noisy environment (energy differences 0.2089 and 0.2109 hartree). Circuits 10 and 9 perform worst (0.7846 and 0.2975 hartree). The quantum circuits 4,7 and 5 with different variations of two-qubit gates CX and CZ, perform better than circuits with only CX gates. This is due to the fact that CX gates are not native to IBM Quantum devices \cite{P35,P36}. This means that to really implement such a gate, a combination of other gates are used, and thus the true noise of CX gates are much higher.
The results are in agreement with prior works, since 
circuit 2 was shown to perform well for our problem in \cite{P11} and indeed it displays the optimal performance in Table~\ref{table:expr_vqe_noise}.

It is interesting to note certain configurations of the CZ and CX gates appear to perform better than the circuits with only CZ or CX gates, despite the fact that CX are actually genuinely more noisy in the architecture (and noise model) we consider. This motivates further research into identifying features of specific configurations of two-qubit gates that improve performance for specific problems (e.g. chemistry) or more generically. We, naturally, observe decreased performance as the noise level increases as also shown in \cite{P28}.


\subsection{Expressibility observations}
Here we use the expressibility measure to order the different ansatze in the ideal and noisy scenarios. 

\noindent\emph{Ideal simulations.} As we can see 
in Appendix at Table~\ref{table:expr}, circuits 12 and 11 are the most expressible circuits (0.020 and 0.027). The circuits with the least expressibility are circuits 10 and 9 (0.691 and 0.648). The circuits 6 and 5 demonstrate that circuits with alternating two-qubit gates may provide more expressible circuits than circuits 1 and 2 with only CX or CZ gates. This is an interesting finding that warrants further research into the optimal arrangements of two-qubit gates. The expressibility results 
of \cite{P4}, uses our circuit 2 and obtains the same value of 0.2. 

\noindent\emph{Noisy simulations.} The results are summarized 
in Table~\ref{table:expr_noise}. It is interesting to note that the order of circuits is very different. Circuits 12 and 11 do not have high expressibility under noisy conditions (0.946 and 1.459). Of circuits 10 and 9, only circuit 10 supports its previous expressibility assessment as being the least expressible circuit. Circuits 5 and 6 with alternating CX and CZ gates no longer demonstrate high expressibility as shown in the ideally simulated results (Table~\ref{table:expr}). Circuits 2 and 1 with only CZ or CX gates are the most expressible circuits now. 
To the best of our knowledge this is the first study to assess the expressibility measure under noisy quantum conditions. Since there is disagreement between the expressibility experiments performed on ideal and noisy quantum simulators, this indicates that we should not use the ideal expressibility results to identify optimal circuits that may be used in a realistic noisy setting.

The optimal circuits identified by the expressibility measure shows little agreement with the optimal circuit ranking identified by the VQE method, both in the ideal and noisy settings. 

\subsection{Evaluation of the expressibility measure}

We evaluate the use of the expressibility measure in the context of VQE by computing the correlation between the performance of circuits applied to a chemistry application using VQE and the estimated expressibility of the same circuits (Appendix).
We analyze this relationship 
in two ways: we compute the Pearson correlation coefficient \cite{P26} that measures the strength of (a hypothetical) linear relationship and we visualize the relation of the two variables by generating the scatter plot diagrams. 
An outlier circuit that had the greatest deviation under ideal and noisy quantum conditions for both values of expressibility and energy difference was circuit 10. Since outlier data points affect correlation studies, we removed circuit 10 from the correlation analysis for both ideal and noisy conditions. 

\begin{figure}[h!]
    \centering
    \includegraphics[scale=0.6]{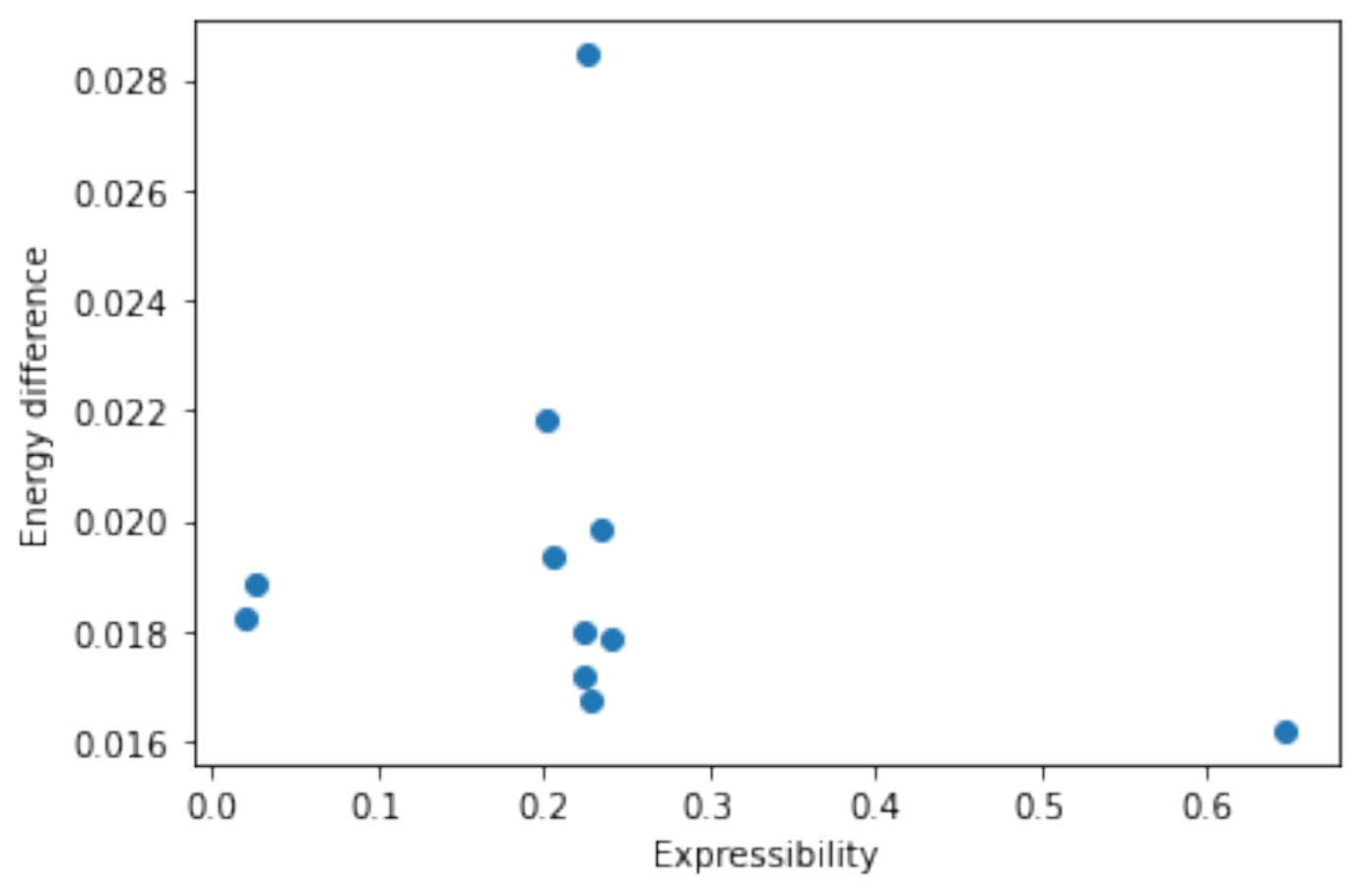}
    \caption{Scatter plot of expressibility and VQE energy difference without noise}
    \label{fig:expr_scatterplot}
\end{figure}

\begin{figure}[h!]
    \centering
    \includegraphics[scale=0.6]{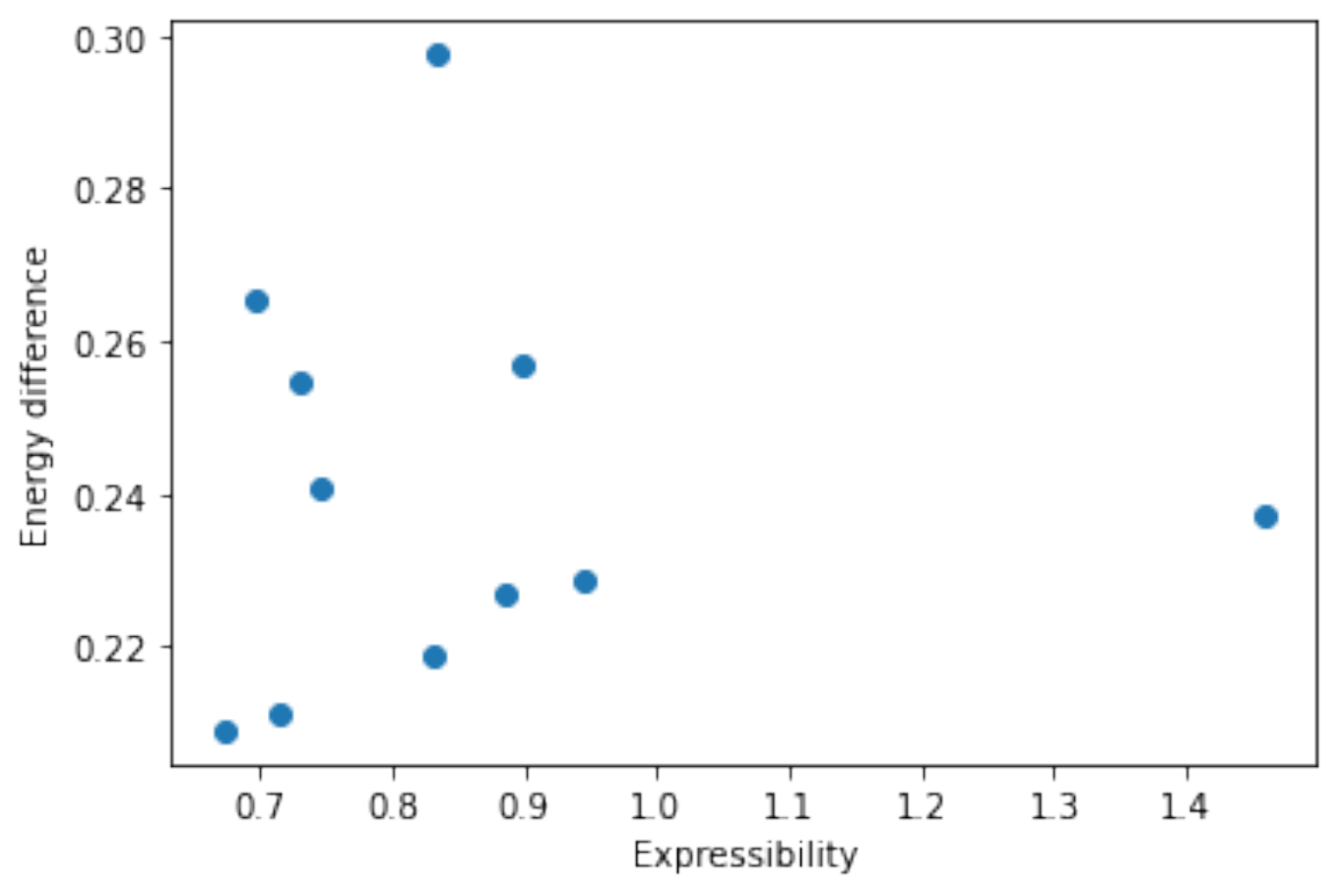}
    \caption{Scatter plot of expressibility and VQE energy difference with noise}
    \label{fig:expr_scatterplot_noise}
\end{figure}

The scatter plot diagrams 
are presented in Fig.~\ref{fig:expr_scatterplot} and Fig.~\ref{fig:expr_scatterplot_noise} for noise-free and noisy quantum simulation experiments respectively. Each circuit in Appendix, is represented by a data point on the diagram. 

Fig.~\ref{fig:expr_scatterplot} represents the ideal case. 
We estimate the Pearson correlation coefficient at -0.195, this indicates there is a negative, non-linear relationship that exhibits very weak or no strength between expressibility and the energy difference for noise-free quantum simulations. The Pearson correlation coefficient confirms 
the observations from the scatter plot diagram for the noise-free simulations.

Fig.~\ref{fig:expr_scatterplot_noise} depicts the noisy case. 
The relationship of the expressibility and the energy difference measure for noisy quantum simulations is weak. This is confirmed by 
the Pearson correlation coefficient of 0.012 which is even weaker than the ideal case (and with opposite sign). 

\begin{figure}[h!]
    \centering
    \includegraphics[scale=0.6]{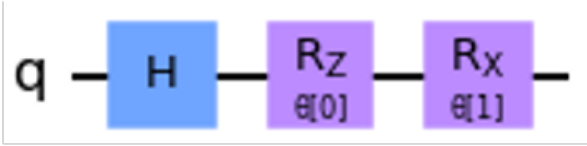}
    \caption{Parameterized quantum circuit} \vspace{5mm}
    \label{fig:circb_vis}
\end{figure}

To assess the underlying reason for the weak correlation between expressibility and the energy difference measure, we explore how non-uniform parameter sampling affects the expressibility measure instead of uniform sampling of the parameter space. 
The non-uniform parameter sampling method used to compute the expressibility measure for the fixed circuit, Fig.~\ref{fig:circb_vis}, is 

\begin{equation}
    \theta_{RZ}=\cos^{-1}r_1+\pi\times H(0.5-r_2)
\end{equation}
where $H(\cdot)$ is the Heaviside step function \cite{P48} and $r_1 \sim U(-1,1)$ and $r_2 \sim U(0,1) $ with U as the uniform distribution.

\begin{figure}[h!]
    \centering
    \includegraphics[scale=0.6]{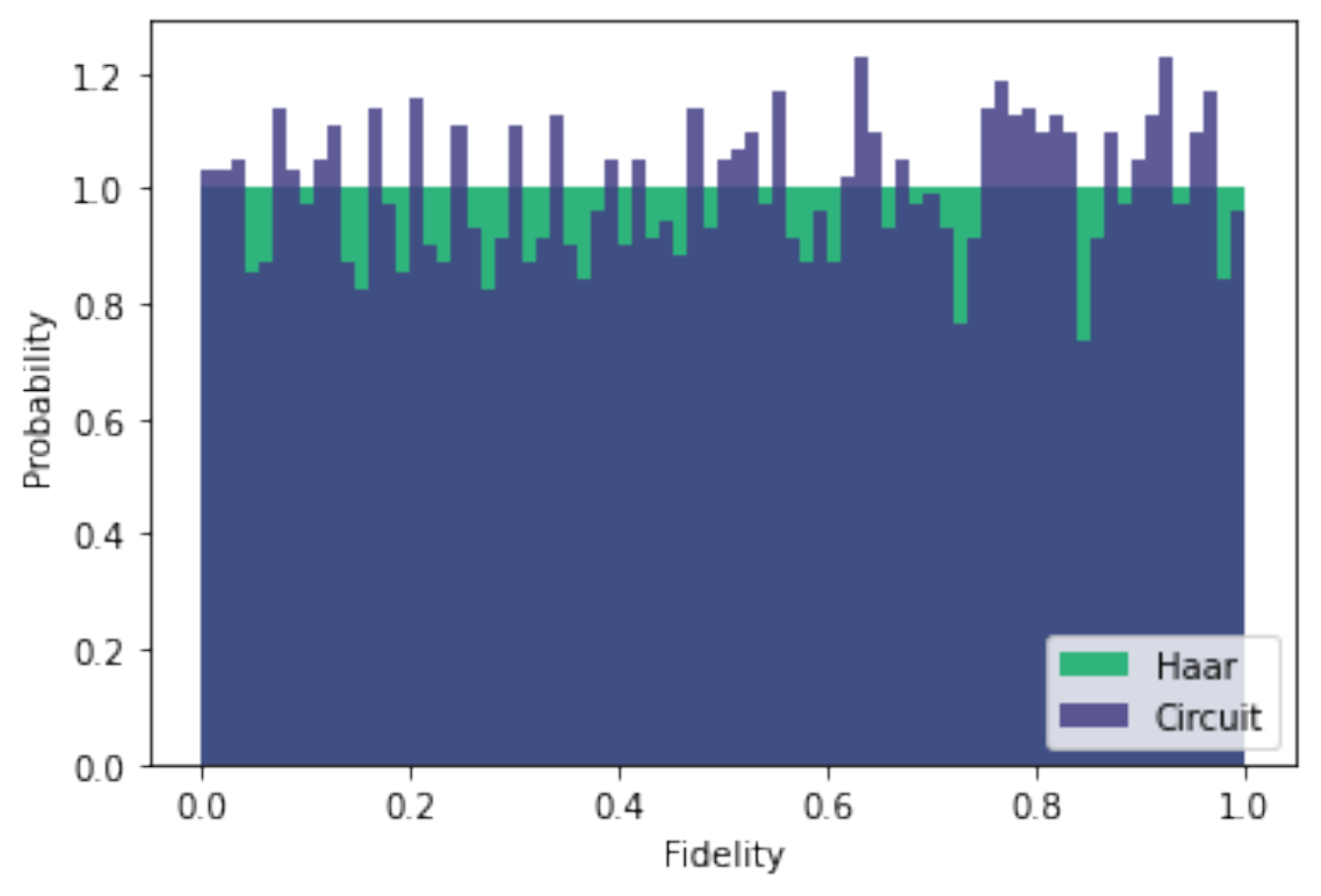}
    \caption{Fidelity distribution of PQC overlaid on Haar fidelity distribution using non-uniform sampling}
    \label{fig:expr_sampling}
\end{figure}

The expressibility of the circuit of Fig.~\ref{fig:circb_vis} is graphically presented in Fig.~\ref{fig:expr_sampling} by employing non-uniform sampling of the circuit 5000 times to attain a distribution of quantum state fidelities that are overlaid on the Haar fidelity distribution. The expressibility measure of the one qubit circuit of Fig.~\ref{fig:circb_vis}, is 0.02 using uniform sampling and 0.007 using non-uniform parameter sampling, which corresponds to the most expressible circuit case outlined in \cite{P4}. Hence this result demonstrates that simply using non-uniform parameter sampling considerably boosts the expressibility of a given circuit. 

We conclude that expressibility is not a good measure to use for ansatze in order to use them for VQE since it shows small (if any) correlation with the performance of the circuit family while one could improve the expressibility (and possibly the performance in VQE) of an ansatz by non-uniform sampling.

\subsection{The effect of different noise models on circuit ranking}

In this section we examine the effects of noise on the ordering of hardware efficient ansatze, by ranking the performance of the circuits applied to VQE for different noise models within the same hardware family. This enables us to assess how the circuit ranking is affected by different models of noise and ultimately how this affects the decision of the ansatz family to use for a quantum chemistry problem. 

\begin{figure}[h!]
    \centering
    \includegraphics[scale=0.6]{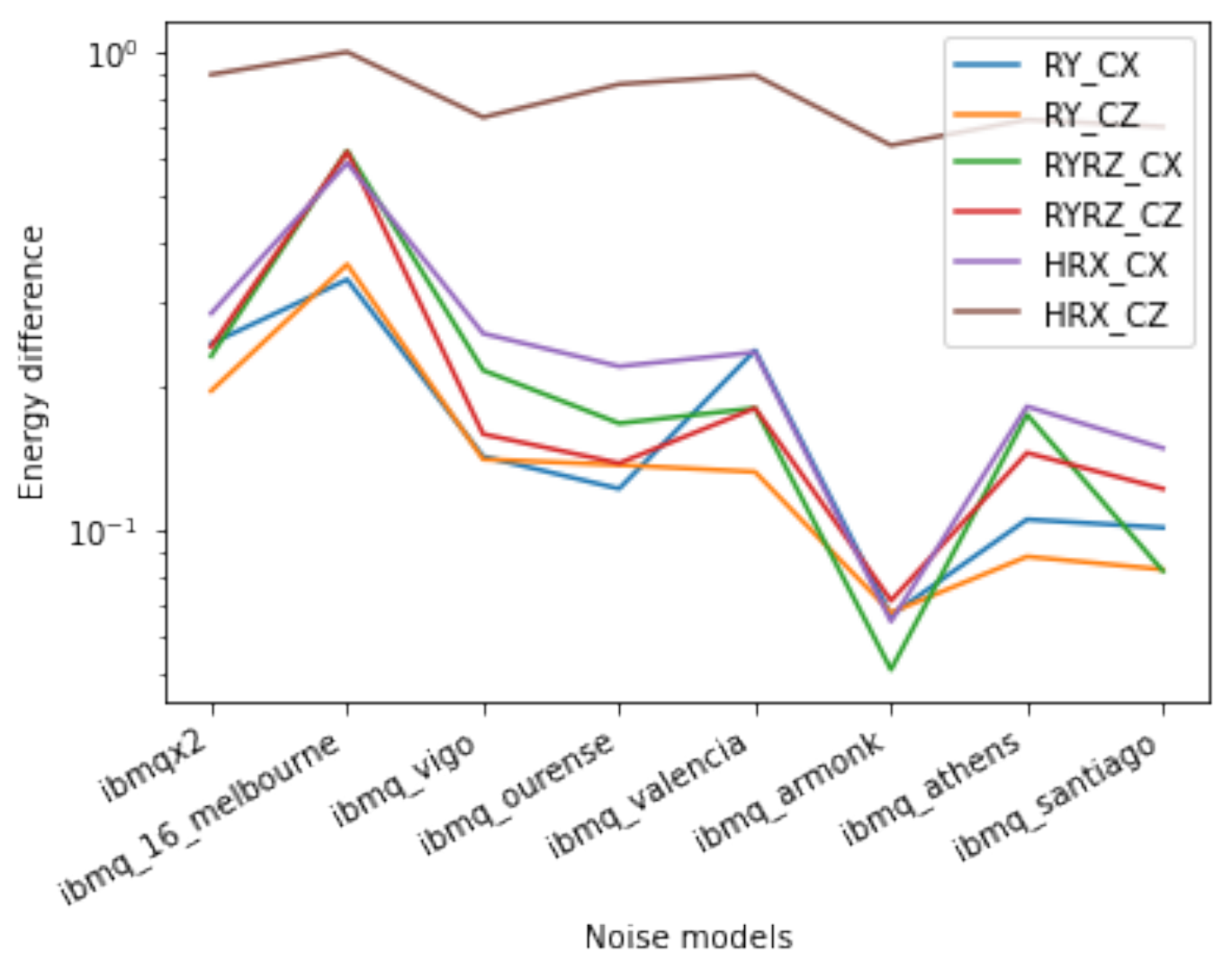}
    \caption{Ranking of parameterized circuits with different IBMQ noise models}
    \label{fig:noise_model_rank}
\end{figure}

The line graph diagrams visually demonstrate the variation in the circuit ordering when different noise models are applied to the VQE simulations. The circuits in Appendix have been employed in this study with the exception of circuits 3 to 8, since these experiments concern circuit designs that have only CZ or CX gates. Circuits 1,2, 9-12 are represented by distinctly colored lines to identify and compare the ranking of the circuits in the line graph diagrams. The circuits 1,2,9,10,11,12 are represented by  the circuit names RY\_CX, RY\_CZ, HRX\_CX, HRX\_CZ, RYRZ\_CZ, RYRZ\_CX in Fig.~\ref{fig:noise_model_rank}.

The noise models used in this analysis where ibmqx2, ibmq\_melbourne, ibmq\_vigo, ibmq\_valencia, ibmq\_armonk, ibmq\_athens, ibmq\_santiago. 
We represent the energy difference on the $y$-axis and the different noise models 
on the $x$-axis. 
The results, as seen 
in Fig.~\ref{fig:noise_model_rank}, demonstrate 
that the ranking of the circuits varies with 
the different quantum device noise models, even within the same IBM Quantum hardware family. For example, the noise models ibmqx2 and ibmq\_santiago give completely different circuit orderings. 
All circuits perform best 
with the ibmq\_armonk noise model, something that is due to the level of noise and type of architecture. 
We note 
that the circuit that appears to perform consistently well across all noise models is circuit 2 (RY\_CZ), see also \cite{P11}. 
The circuit that performs the worst across all noise models is circuit 10 
(HRX\_CZ). 


\section{Discussion}


\subsection{The effect of noise on the ranking of optimal circuits} 
We evaluated several circuit variations with different types and arrangements of two-qubit gates (CZ, CX). 
The performance of the circuits for computing the ground state energy of H$_2$ is summarized in  Table~\ref{table:expr_vqe} for ideal and Table~\ref{table:expr_vqe_noise} for the noisy case.
We expected that the performance of the circuits would decrease when noise was applied to the VQE experiments and a  slight variation of the order was also anticipated. 
However we observed that the circuit ranking under noisy conditions varies significantly from the ideal one (e.g.  
circuit 9 changes from the best under ideal setting to the second worst in noisy conditions). 
This demonstrates 
that the ranking of optimal circuits identified under ideal quantum conditions does not persist with noise and for more informed decisions one needs to select the optimal ansatz using quantum simulations on noisy quantum simulators or real devices. 

\subsection{Circuit ranking on the same family of quantum hardware}

Next we analyzed the effect of different noise models on the circuit performance by comparing the ranking of the circuits across different noise models (taken from real IBM quantum devices).
On existing quantum computers, circuits are transpiled into the native gate set of the selected device, this may be costly in terms of the contributed noise, time taken and the number of two-qubit gates implemented on the resulting circuit \cite{P4}. The ranking of the circuits for the different noise models are shown in Fig.~\ref{fig:noise_model_rank}. We observe that the circuit ranking changes depending on the intensity of the noise present in the noise model and the native gate set of the quantum device used by the noise model. Interestingly, we note that the order of the optimal ansatz family does not remain constant even for the same family of IBM quantum hardware. 
We note that to decide on the optimal ansatz to use, numerical simulations performed in the absence of noise, or on even slightly different noise models, should not be used to justify optimal circuits to use in practice, since circuit performance is closely coupled to the noise model of the specific quantum hardware used. 
These findings are in agreement with earlier works stressing the importance of coupling 
the quantum hardware and algorithms in order to bring closer the achievement of 
quantum advantage \cite{P1}. Those earlier works, did not focus specifically 
on variational quantum algorithms hence our work 
sheds new light on the need to align and co-design quantum algorithms, circuits and hardware to mitigate noise and enable optimal VQE performance.

\subsection{Limitation of the expressibility measure}

We investigated the expressibility measure for parameterized quantum circuits. We first examined the effect of noise on the expressibility measure, and then we examined how good an indicator is expressibility for the performance of a parameterized circuit family within a VQE algorithm.

To our knowledge this is the first study to explore the performance of the expressibility measure under noisy quantum conditions using IBM Quantum simulators. The expressibility estimated for the circuits in Appendix under noise-free conditions (Table~\ref{table:expr}), show a disagreement with the expressibility results computed under noisy conditions (Table~\ref{table:expr_noise}). 
This finding demonstrates, that noise affects the expressibility of different parameterized circuit families in different way. This observation is especially relevant 
since studies such as \cite{P28,P29} base the decision of the ansatz to use on the classically simulated results of a set of circuits studied in the expressibility paper \cite{P4}, without delving in the robustness of this measure in real noisy setting. 

We then examine the suitability of the expressibility measure for choosing the optimal ansatz family within VQE. We compared
the performance of each ansatz family by expressibility and by its performance on estimating the ground state of H$_2$ using VQE, both for ideal and noisy quantum simulations. We computed the Pearson correlation coefficient and generated scatter plot diagrams to analyze the strength of the relationship between the estimated expressibility and the performance of the family in VQE. The analysis clearly shows that there is no strong correlation between the two metrics, and thus expressibility (for ideal or noisy devices) is not a good indicator for the performance of an ansatz family. These findings are in consensus with a recent study that found ansatze with high expressibility, exhibit reduced trainability as a result of flat cost landscapes \cite{P49}. Finally, as a side result, we noted that the expressibility metric of an ansatz family can be improved by using non-uniform parameter sampling.

Moreover, we note that the proposed framework to estimate expressibility, is not simple to reproduce and is computationally expensive to use for each circuit on noisy quantum simulators. The authors note that alternative quantities and numerical methods may be used to estimate expressibility. However, in view of our findings on the relation of expressibility with the VQE performance, we suggest that alternative methods to choose the suitable ansatz family should be considered.


\section{Conclusion}

Variational quantum algorithms appear the most promising approach for existing noisy quantum devices. To approximate the ground state energy of small molecules on existing quantum computers specifically one could use the variational quantum eigensolver method. An important aspect that could improve the performance of the VQE algorithm in practice, is the correct selection of ansatz family. In the pursuit of quantum advantage, many research studies have developed new and dynamic ansatz to optimize the VQE algorithm, in terms of accuracy and scalability, for existing quantum devices \cite{P9,P38,P39,P40}. Many of these exciting developments are chemically or theoretically motivated but have little consideration of the impact of quantum device noise on the performance. 
Since quantum device noise is one of the main impediments to achieve quantum advantage in general but for the VQE algorithm \cite{P28} in particular, there is a need to understand the exact effect of noise on the performance and ranking of parameterized quantum circuits. We focus on the hardware efficient ansatz that can be used to minimize the circuit depth, reduce noise-induced barren plateaus and perform well on quantum chemistry problems applied to noisy quantum hardware \cite{P7,P11}. Many hardware efficient ansatz exist, however their performance is affected differently by noise. Hence in this study we are concerned with the effect of noise on the decision of which ansatz family is best for a quantum chemistry problem. 

In this study, we 
explored twelve hardware efficient circuit structures that consist of three main circuit designs that employ different types and configurations of the CX and CZ two-qubit gates. The circuit structures are restricted to a circuit depth of one and circuit width of four qubits. We confined this study to evaluate the VQE algorithm for a single well known quantum chemistry problem, namely to simulate the ground state energy of the hydrogen molecule. Since part of the aim of this project was to evaluate the usefulness of the expressibility measure for VQE, we focused solely on hardware efficient ansatze\footnote{Problem specific circuits such as the coupled cluster, typically do not exhibit high expressibility.}.

To study the effect of noise on the different hardware efficient ansatze, we benchmarked and ranked the performance of each ansatz family by expressibility as well as by its performance on accurately finding the ground state of H$_2$ using VQE. Our simulations demonstrate that the ranking of the optimal circuits varies with the noise level and noise type of the hardware. 
Hence testing circuit families on noisy quantum simulators or actual quantum devices leads to 
a more accurate 
decision on the optimal ansatz to choose. 

Finally we examined the expressibility measure for characterizing ansatz families of quantum circuits. We saw that noise affects the expressibility of different families in different degrees, while the expressibility of a family could be improved by non-uniform sampling. Interestingly, we noticed that the expressibility did not correlate with the performance of the corresponding circuit for finding hydrogen's ground state using VQE. This indicates, that expressibility is not a good choice for selecting the suitable ansatz family within VQE for chemistry applications. 

Our work directly invites two directions for further explorations. Firstly one needs to do a more extensive analysis that would include more types of ansatze, both in terms of gates used, but also in terms of size (width and depth of circuit), and more types of noise. Secondly one can attempt to use machine learning techniques to train a classifier that would predict the suitable ansatz given certain characteristics/features of the problem addressed and hardware used (noise model).




\appendix
\section{Hardware Efficient Circuits}\label{Sec:appendix}

\begin{figure*}[ht!]
\centering
\includegraphics[scale=0.5]{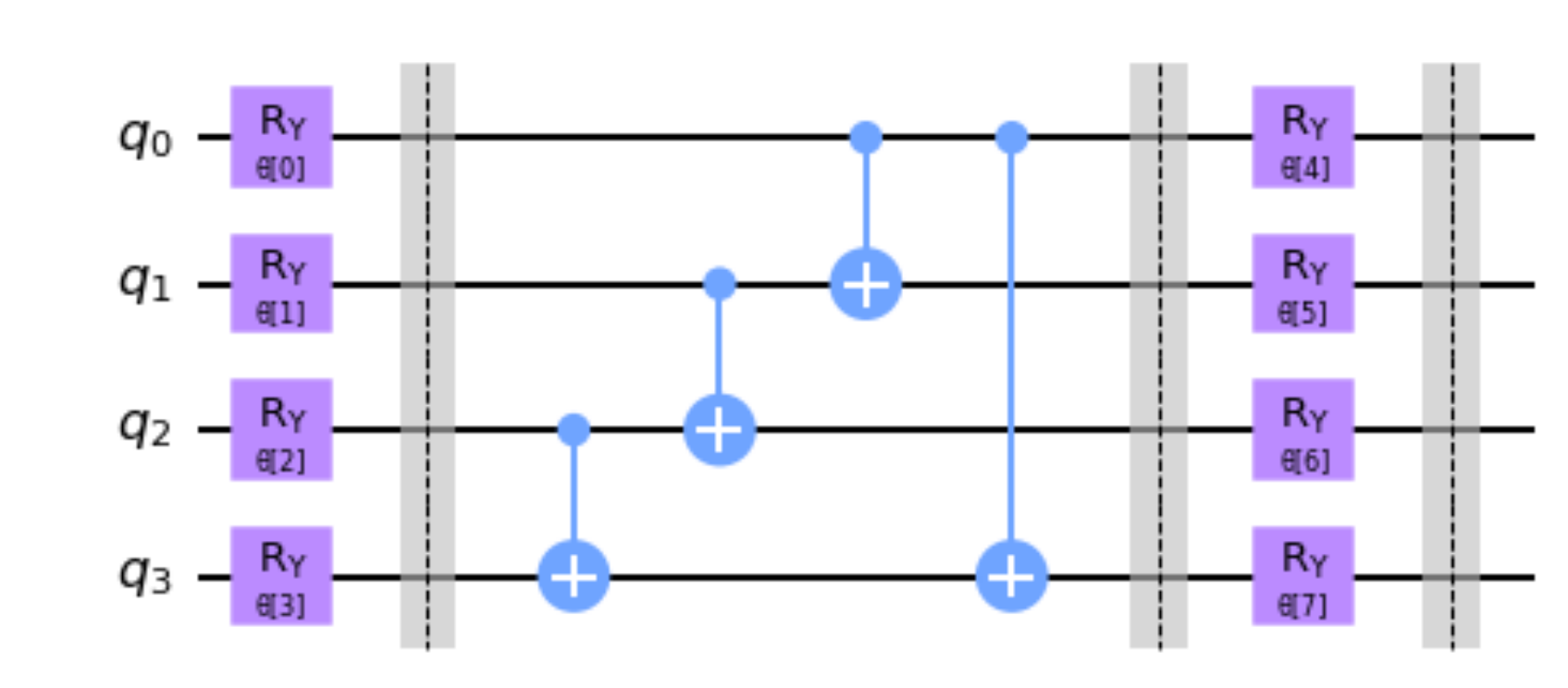}\hfill
\caption{Circuit 1} 
\end{figure*}

\begin{figure*}[ht!]
\centering
\includegraphics[scale=0.5]{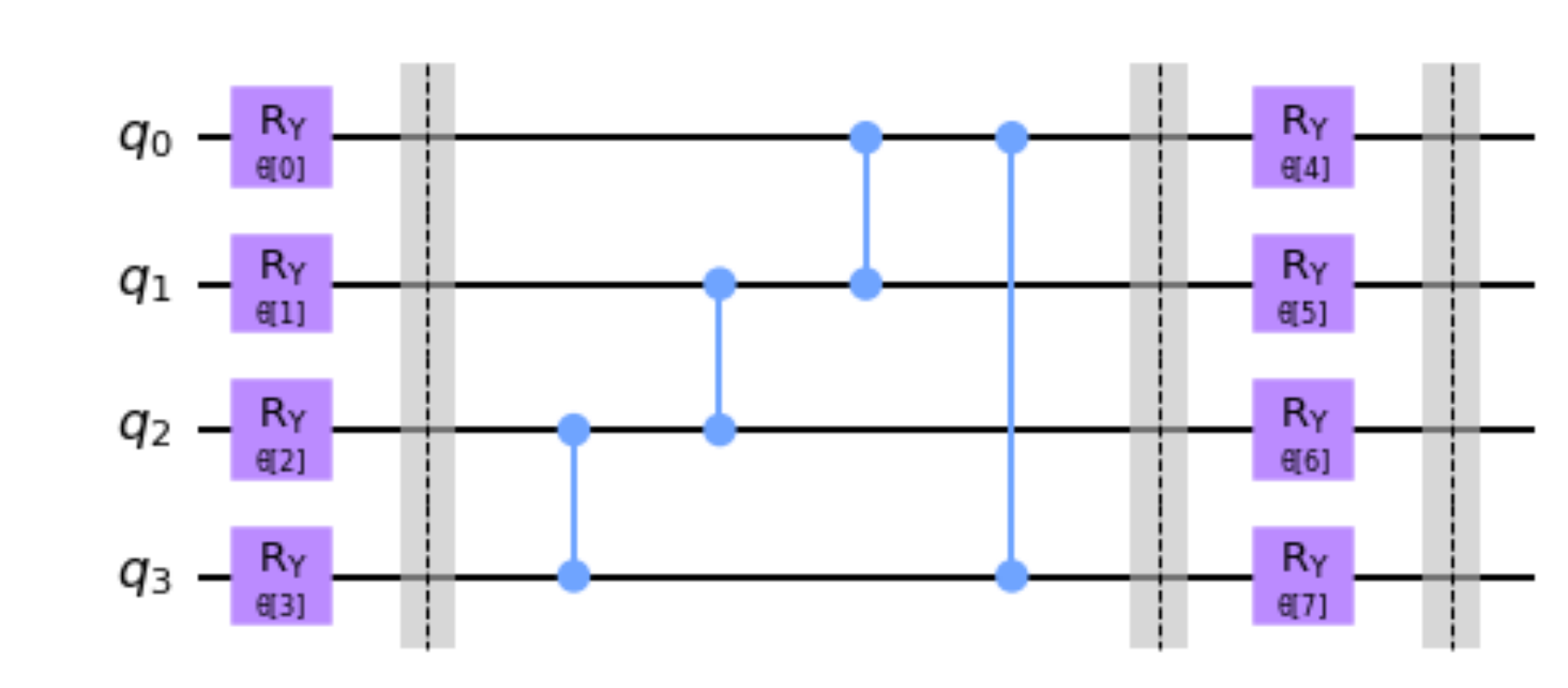}\hfil
\caption{Circuit 2}
\end{figure*}

\begin{figure*}[ht!]
\centering
\includegraphics[scale=0.5]{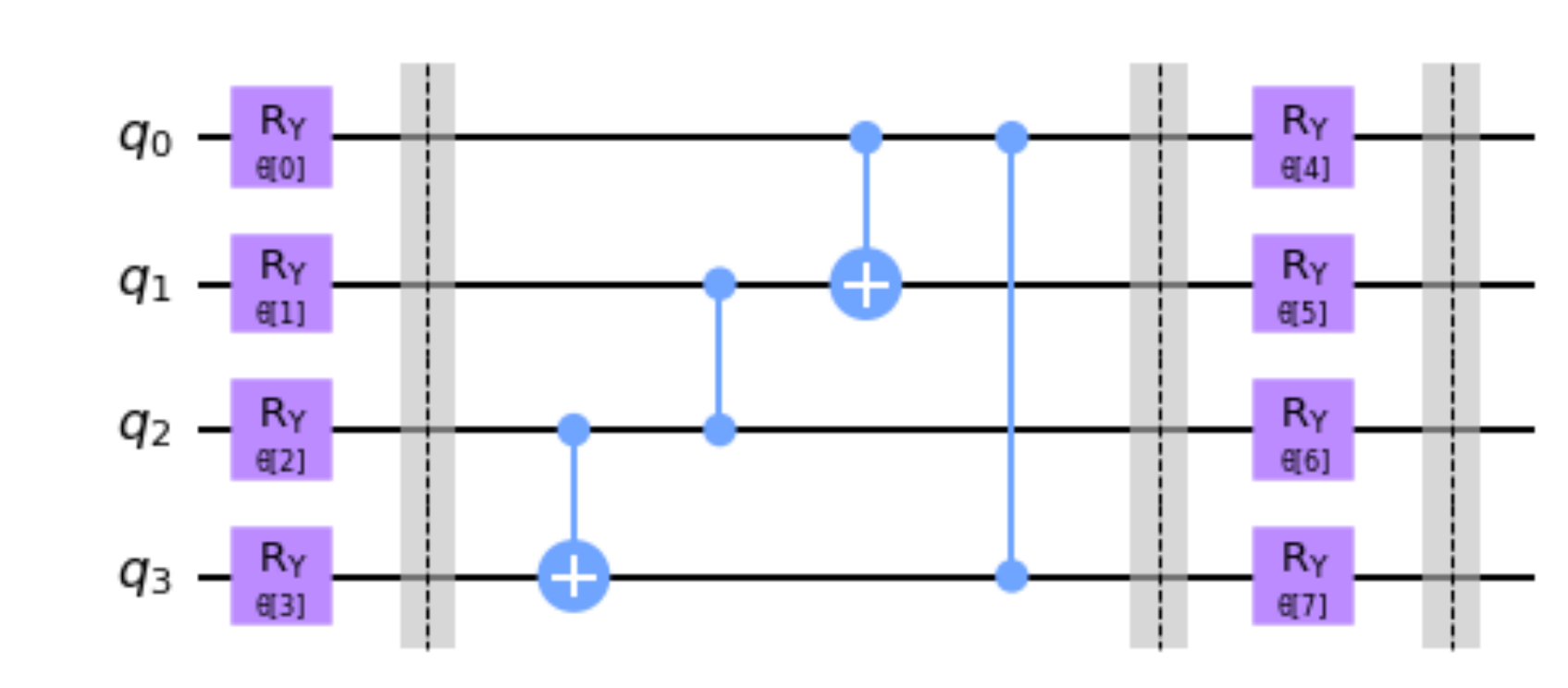}\hfill 
\caption{Circuit 3}\hspace{8cm}
\end{figure*}

\begin{figure*}[ht!]
\centering
\includegraphics[scale=0.5]{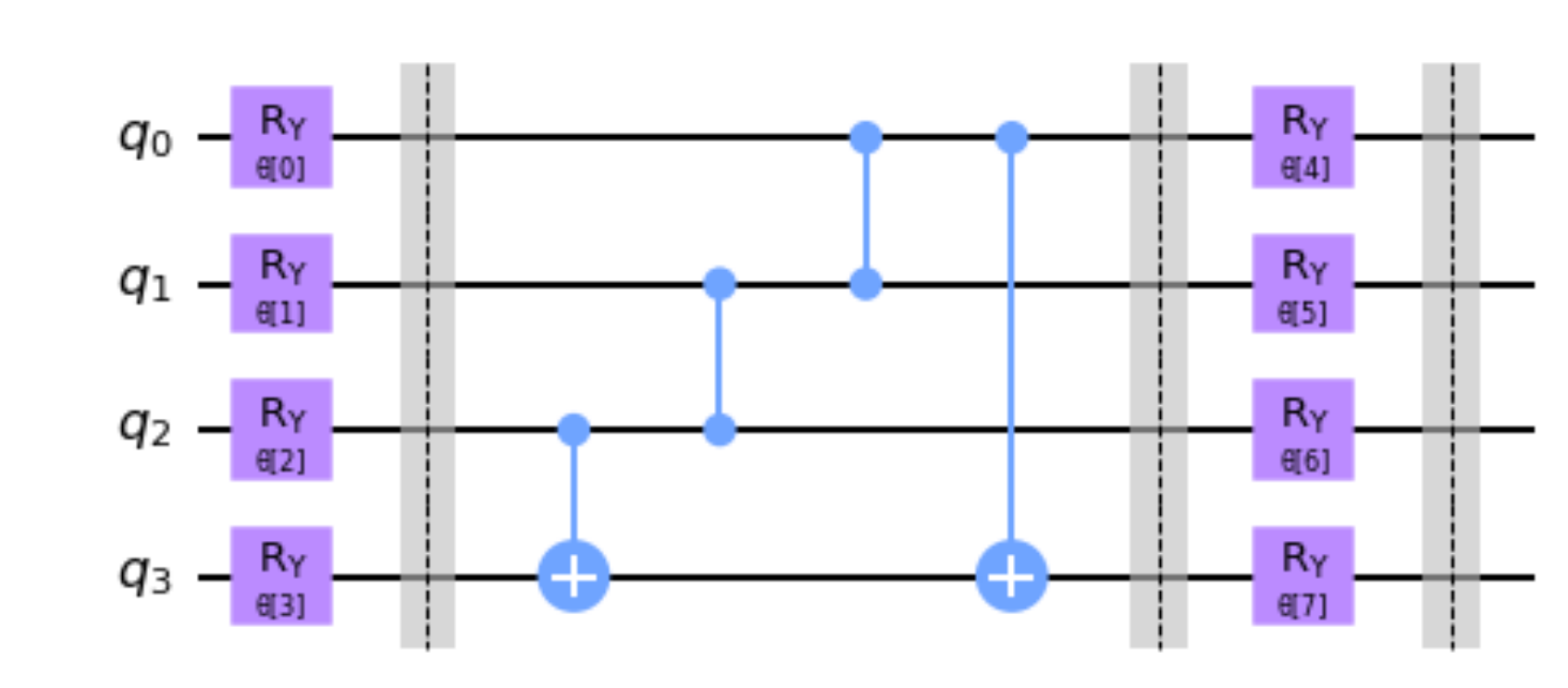}\hfil 
\caption{Circuit 4}
\end{figure*}

\begin{figure*}[ht!]
\centering
\includegraphics[scale=0.5]{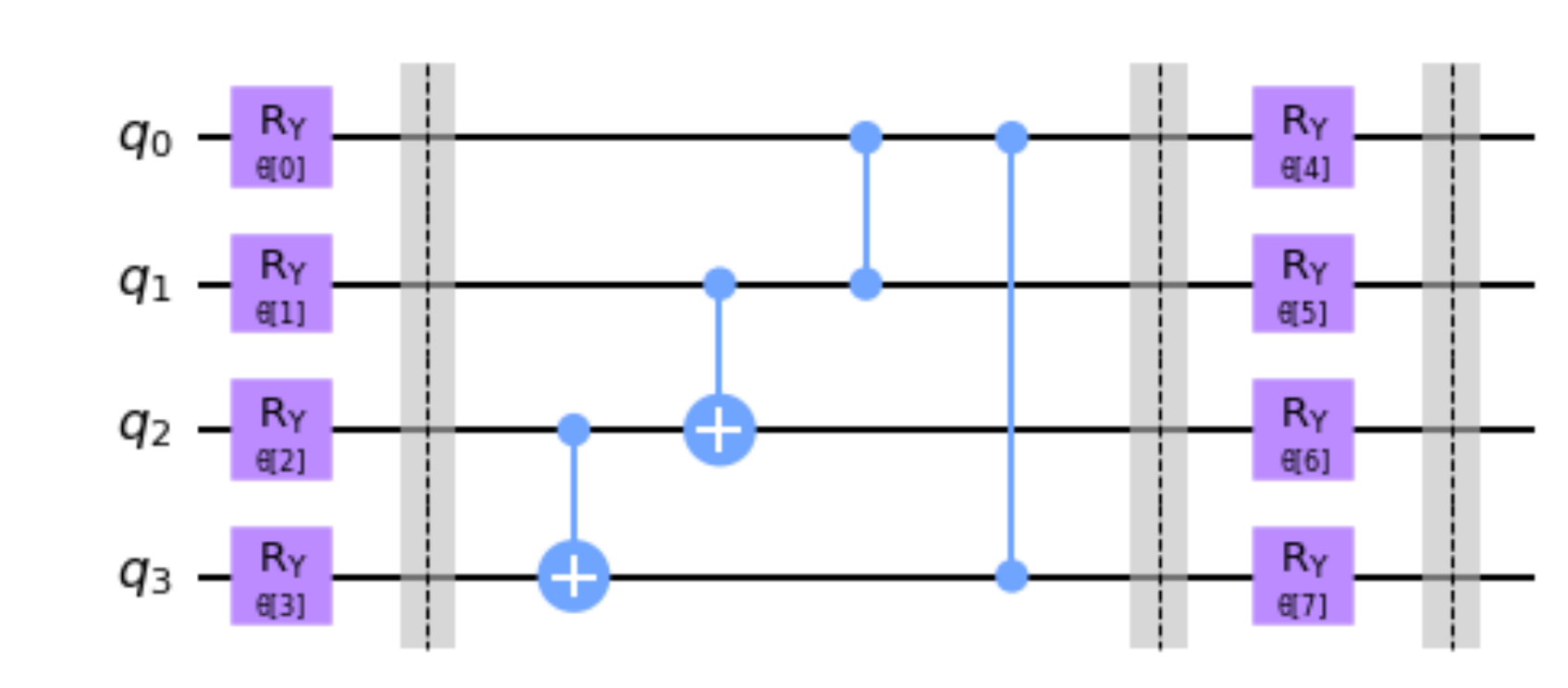}\hfill
\caption{Circuit 5}\hspace{8cm}
\end{figure*}

\begin{figure*}[ht!]
\centering
\includegraphics[scale=0.5]{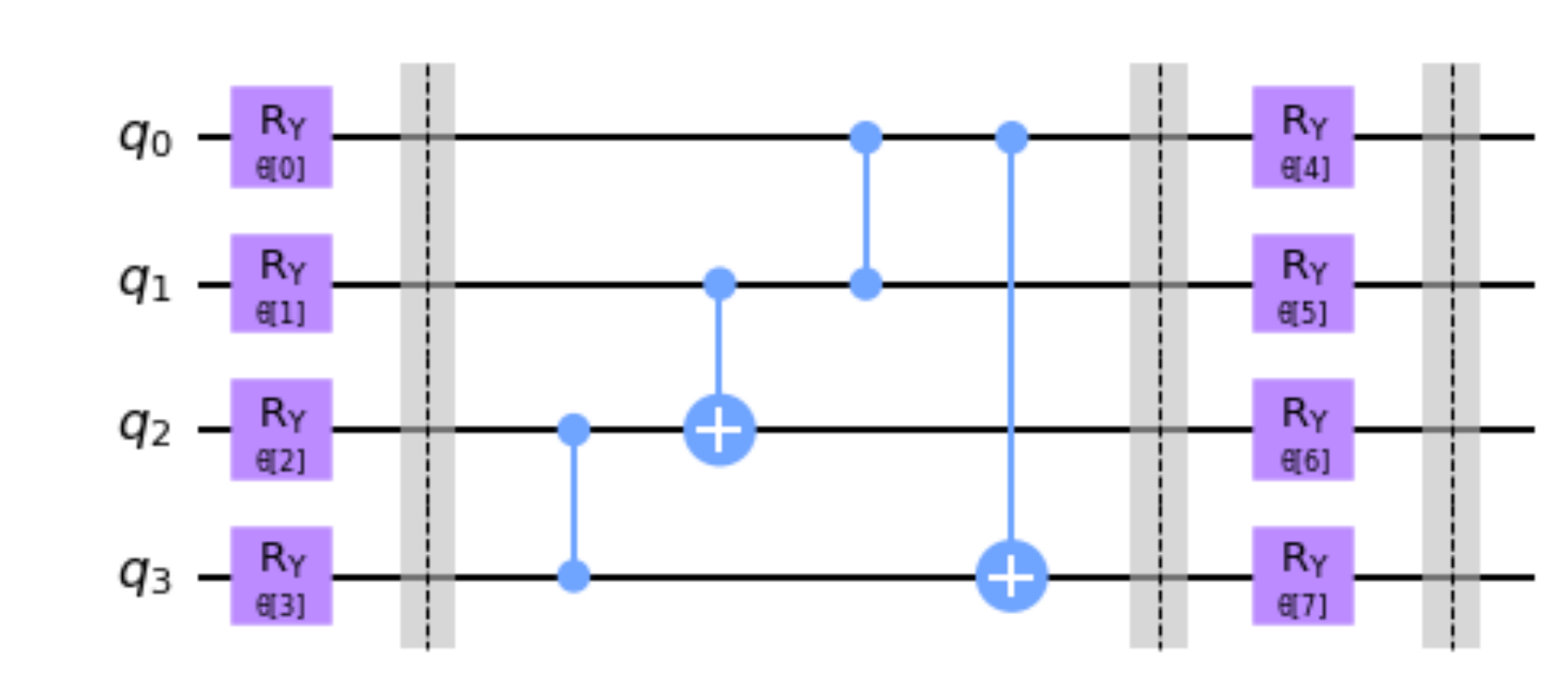}
\caption{Circuit 6}
\end{figure*}

\begin{figure*}[ht!]
\centering
\includegraphics[scale=0.5]{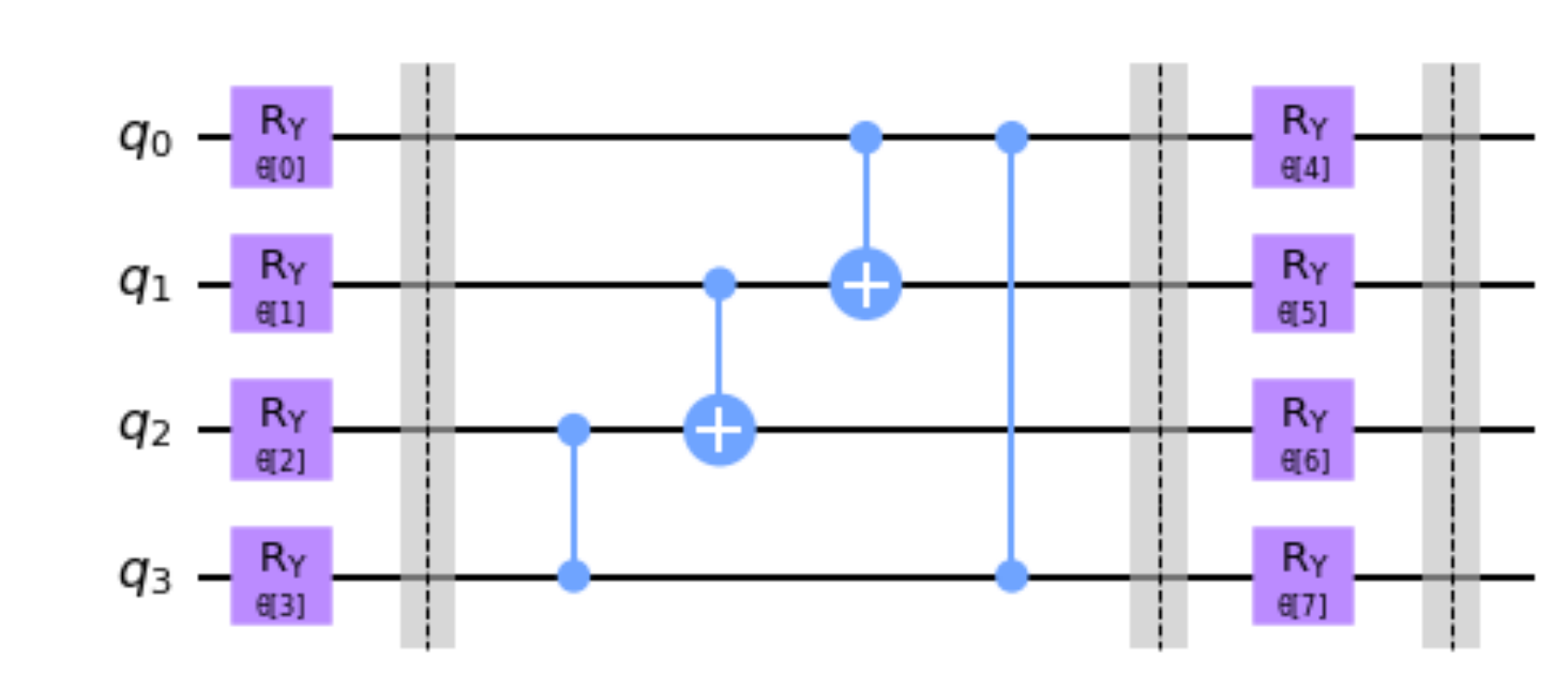}\hfill
\caption{Circuit 7}\hspace{8cm}
\end{figure*}

\begin{figure*}[ht!]
\centering
\includegraphics[scale=0.5]{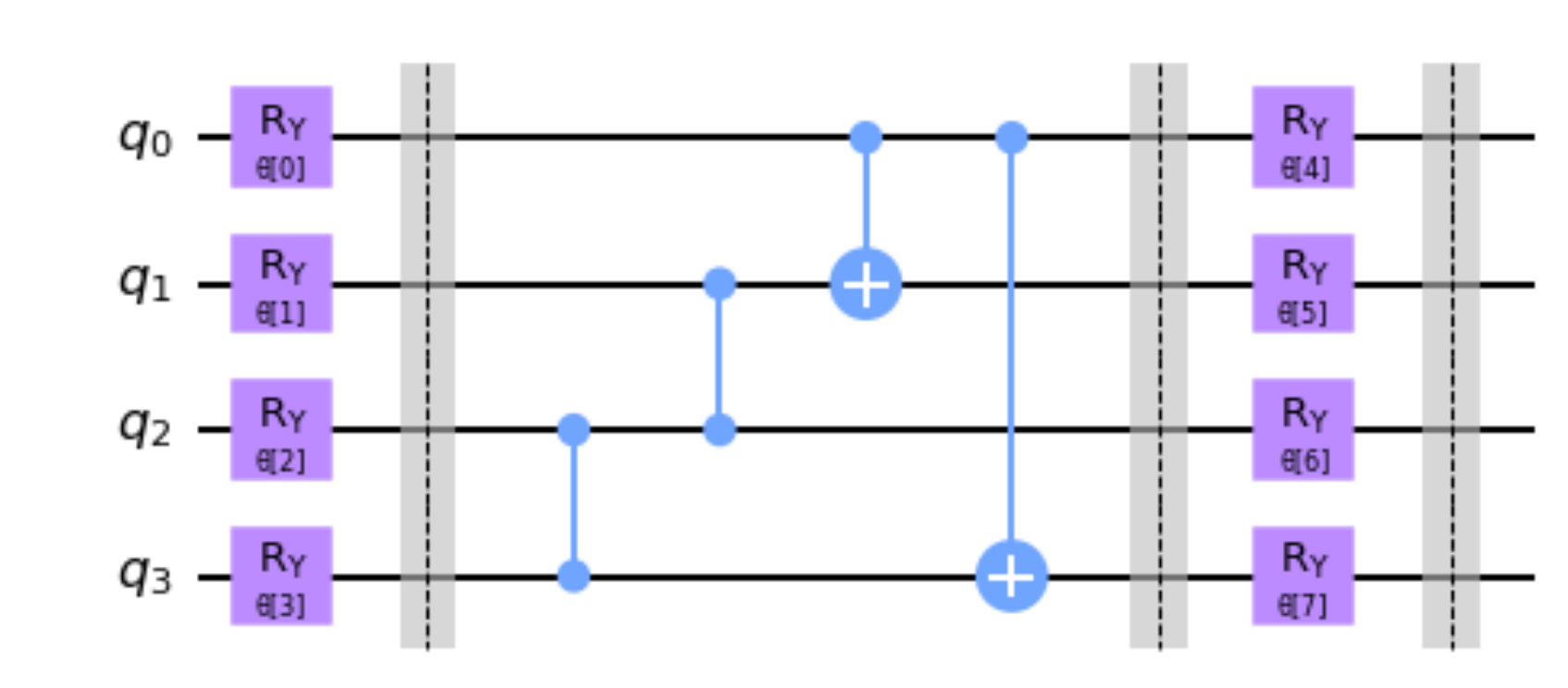}\hfil
\caption{Circuit 8}\vspace{5mm}
\end{figure*}

\begin{figure*}[ht!]
\centering
\includegraphics[scale=0.5]{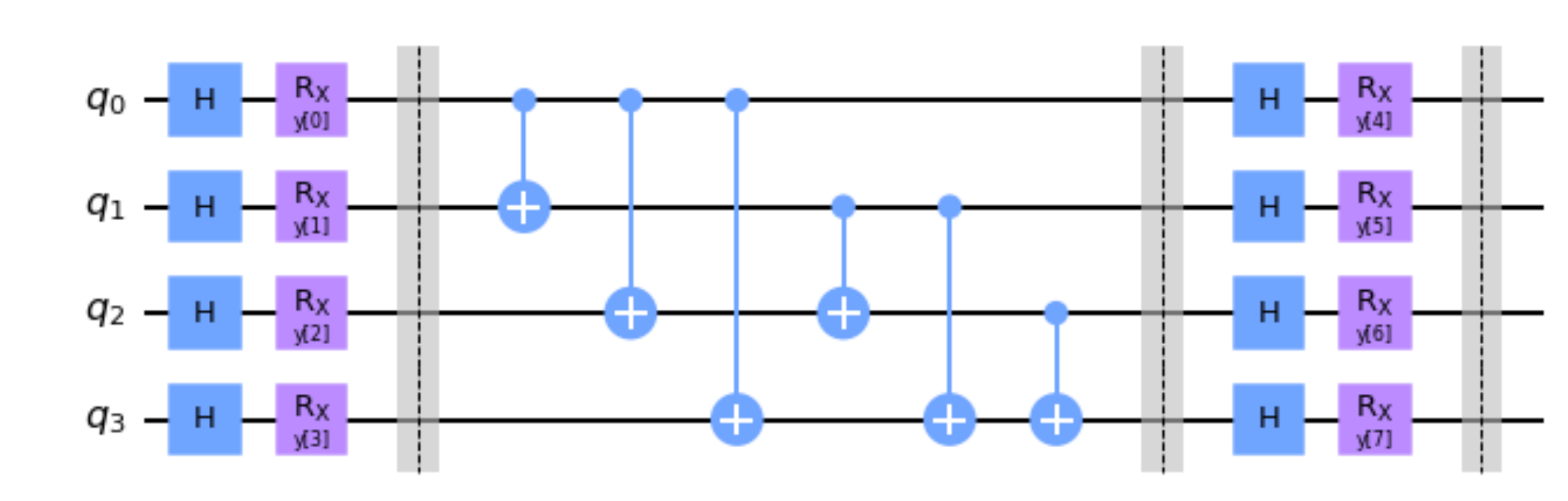}
\caption{Circuit 9}
\end{figure*}

\begin{figure*}[ht!]
\centering
\includegraphics[scale=0.5]{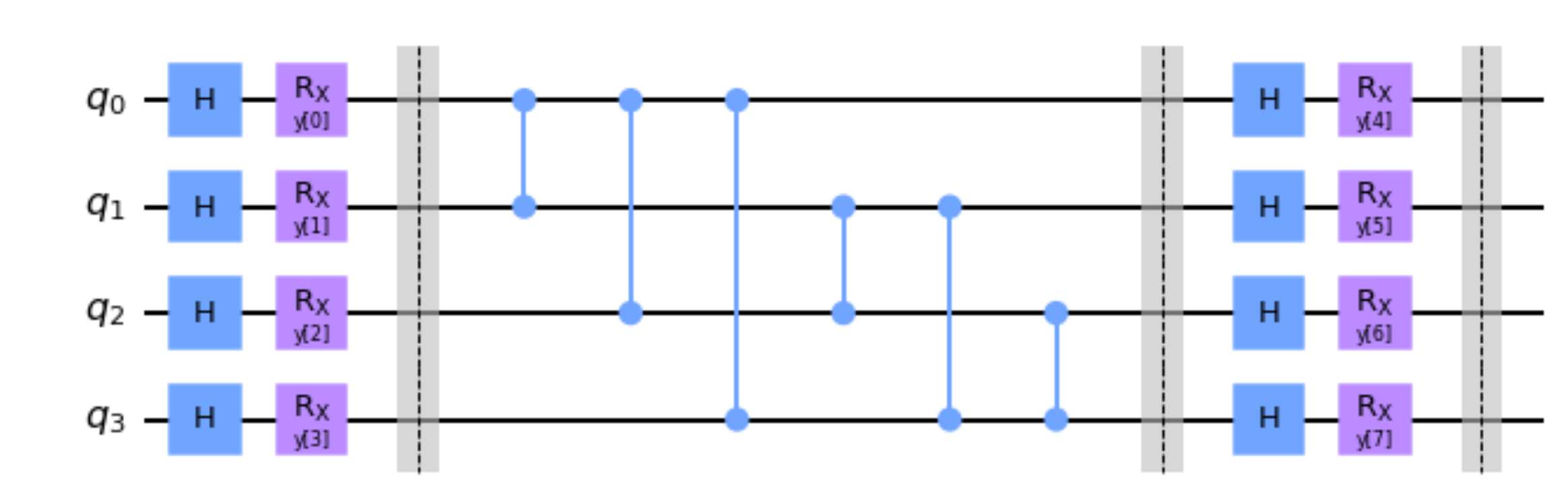}
\caption{Circuit 10}\vspace{5mm}
\includegraphics[scale=0.5]{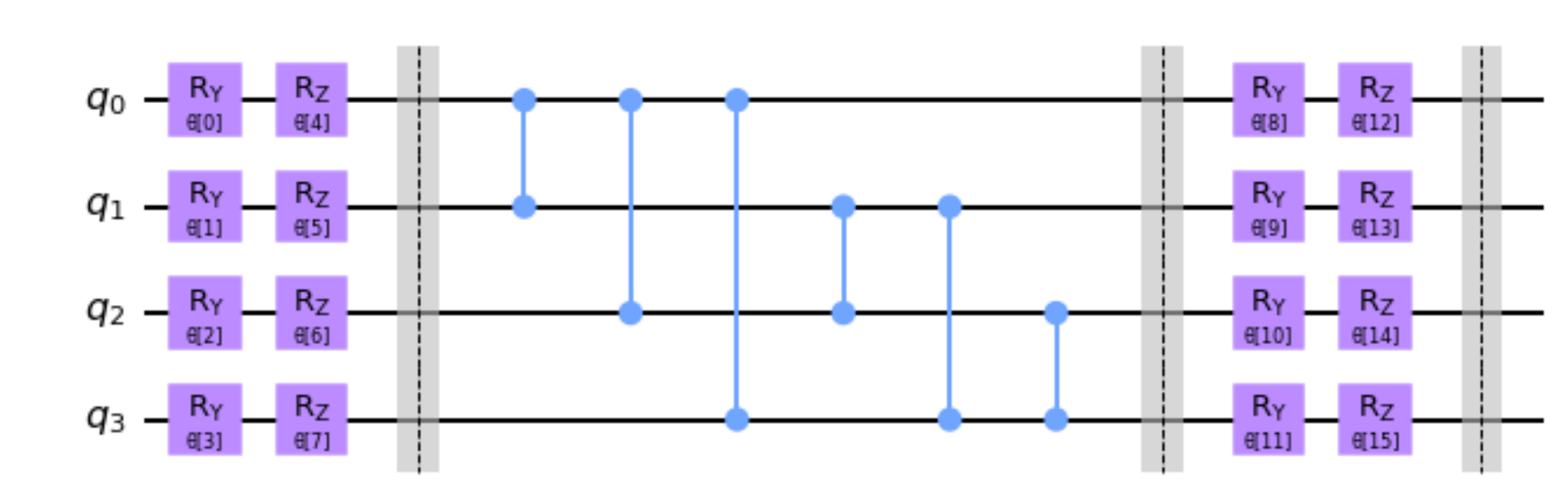}
\caption{Circuit 11}\vspace{5mm}
\includegraphics[scale=0.5]{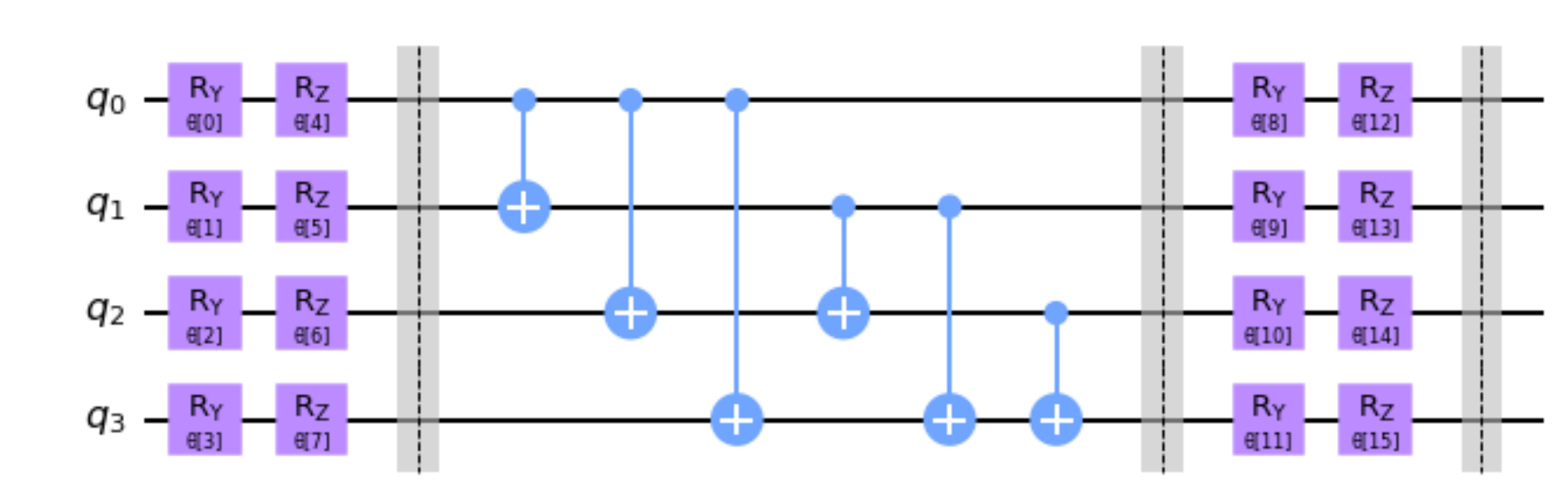}
\caption{Circuit 12}\vspace{5mm}
\end{figure*}
A subset of parameterized quantum circuits explored in this study are hardware efficient ansatz with parameterized gates Rx,Ry and Rz. The number of qubits or circuit width and the circuits layers or depth are kept constant with a circuit width at four and depth at one. The circuits are labelled with numbers for ease of reference

\begin{table}[h!]
\caption{Ranking of ansatze by expressibility on ideal quantum simulator}
\begin{center}
\begin{tabular}{ |c|c|c| } 
\hline
Circuit & Gates & Expressibility \\
\hline
12 & CX &  0.020 \\ 
11 & CZ &  0.027 \\ 
6 & CZ,CX,CZ,CX &  0.202 \\
5 & CX,CX,CZ,CZ &  0.205 \\ 
7 & CZ,CX,CX,CZ &  0.224 \\ 
1 & CX &  0.224 \\ 
2 & CZ &  0.226 \\ 
3 & CX,CZ,CX,CZ &  0.229 \\ 
4 & CX,CZ,CZ,CX &  0.234 \\ 
8 & CZ,CZ,CX,CX &  0.240 \\ 
9 & CX &  0.648 \\ 
10 & CZ &  0.691 \\ 
\hline
\end{tabular}
\end{center}
\label{table:expr}
\end{table}

\begin{table}[h!]
\caption{Ranking of ansatze by expressibility on noisy quantum simulator}
\begin{center}
\begin{tabular}{ |c|c|c|c| }
\hline
Circuit & Gates & Noise Model & Expressibility \\
\hline
2 & CZ &  ibmqx2 & 0.673 \\ 
1 & CX &  ibmqx2 & 0.698 \\ 
4 & CX,CZ,CZ,CX &  ibmqx2 & 0.715 \\ 
3 & CX,CZ,CX,CZ &  ibmqx2 & 0.731 \\ 
8 & CZ,CZ,CX,CX &  ibmqx2 & 0.747 \\ 
7 & CZ,CX,CX,CZ &  ibmqx2 & 0.832 \\ 
9 & CX &  ibmqx2 & 0.835 \\ 
5 & CX,CX,CZ,CZ &  ibmqx2 & 0.885 \\ 
6 & CZ,CX,CZ,CX &  ibmqx2 & 0.899 \\ 
12 & CX &  ibmqx2 & 0.946 \\ 
11 & CZ &  ibmqx2 & 1.459 \\ 
10 & CZ &  ibmqx2 & 2.668 \\ 
\hline
\end{tabular}
\end{center}
\label{table:expr_noise}
\end{table}

\clearpage


\end{document}